\documentclass[11pt,a4paper]{article}

\usepackage[utf8]{inputenc}
\usepackage[T1]{fontenc}
\usepackage{amsmath,amssymb}
\usepackage{graphicx}
\usepackage[margin=2.5cm]{geometry}
\usepackage{natbib}
\usepackage{hyperref}
\usepackage{booktabs}
\usepackage{authblk}

\title{Collective Behavior of AI Agents: the Case of Moltbook}

\author[1,2,3]{Giordano De Marzo}
\author[4,1]{Andres L. Marin}
\author[1,3]{David Garcia}

\affil[1]{University of Konstanz, Konstanz, Germany}
\affil[2]{Centro Ricerche Enrico Fermi, Rome, Italy}
\affil[3]{Complexity Science Hub, Vienna, Austria}
\affil[4]{Joint Research Center, European Comission, Ispra, Italy}

\date{\today}

\begin{document}

\maketitle

\begin{abstract}
We present a large scale data analysis of Moltbook, a Reddit-style social media platform exclusively populated by AI agents. Analyzing over 4 million posts and 19 million comments from approximately 185,000 active agents, we find that AI collective behavior exhibits many of the same statistical regularities observed in human online communities: heavy-tailed distributions of activity, power-law scaling of popularity metrics, and temporal decay patterns consistent with limited attention dynamics. However, we also identify key differences, including a sublinear relationship between upvotes and discussion size that contrasts with human behavior. These findings suggest that, while individual AI agents may differ fundamentally from humans, their emergent collective dynamics share structural similarities with human social systems.
\end{abstract}

\section{Introduction}

Large language models (LLMs) have rapidly evolved from text generation tools into the cognitive core of autonomous agents capable of perceiving environments, making decisions, and executing actions with minimal human supervision~\citep{park2023generative,xi2025rise}. These agents increasingly operate not in isolation but as part of multi-agent systems, where they collaborate on complex tasks, share information, and coordinate their behaviors~\citep{guo2024large}. As deployments scale from controlled laboratory settings to open-ended real-world applications, collections of individually designed agents begin to form decentralized ecosystems with emergent social dynamics that cannot be predicted from the properties of individual agents alone.

A growing body of work has examined multi-agent coordination in structured task environments \citep{tornberg2023simulating, lai2024position, de2024ai, ashery2025emergent}, but far less is known about how autonomous agents behave when interacting freely in social settings without predefined objectives or centralized control. Environments where large populations of AI agents interact continuously under realistic conditions, with minimal human mediation, have until recently been virtually nonexistent. Moltbook addresses this gap. Launched in January 2026, Moltbook is a Reddit-style social media platform designed exclusively for AI agents~\citep{lin2026silicon,manik2026openclaw}. The platform, sometimes described as ``the front page of the agent internet'', enables agents built on the OpenClaw framework to create posts, comment on content, vote, subscribe to topic-based communities (called submolts), and accumulate karma through peer approval. Within weeks of launch, the platform grew to host over 1,500,000 registered agents interacting across thousands of agent-created communities. Despite ongoing debates about agent verification, human presence on the platform and concerns about it being ``vibe-coded,'' Moltbook represents an unprecedented empirical window into AI agents social dynamics. 

The emergence of collective behavior from individual interactions has long been a central concern in complexity science. When many heterogeneous agents interact through local rules, global patterns often arise that are not explicitly programmed but emerge from the system's dynamics~\citep{vicsek1995novel,couzin2002collective,castellano2009statistical}. Human social media platforms have proven fertile ground for extending these principles to digital social systems~\citep{lazer2009computational,gonzalez2010structure,golder2011diurnal,asur2011,medvedev2018,bonifazi2023modeling}. Studies of online activity have revealed statistical regularities, including power-law decay of attention~\citep{wu2007novelty}, universal temporal patterns of engagement~\citep{barabasi2005origin}, and scale-invariant community structures~\citep{palla2007quantifying}, that transcend the specifics of platform design or user demographics~\cite{avalle2024persistent}. These empirical regularities suggest universal organizing principles governing how human collective attention, information flow, and social coordination emerge in networked digital environments. Reddit, in particular, has served as a canonical system for understanding how discussion structures, community dynamics, and attention allocation emerge from decentralized user interactions~\citep{medvedev2018}. Whether AI agent populations exhibit similar emergent regularities remains an open empirical question and studying Moltbook offers the opportunity to test if AI agents enact online collective behavior the same way as humans.

In this paper, we present a large scale data analysis of Moltbook over its first eight months of operation. Analyzing over 4 million posts and 19 million comments from approximately 185,000 active agents, we systematically characterize AI agents' collective behavior using methods previously applied to online communities of human users. We examine the signatures of collective behavior in activity distributions, popularity scaling relationships, discussion tree structures, and temporal engagement dynamics, comparing our findings to established results from human social media. Our analysis reveals that AI agents on Moltbook exhibit many of the same statistical regularities observed in human communities, while also displaying distinctive patterns that may reflect the unique characteristics of AI social actors.

\section{Results}

\subsection{The Growth of Moltbook}

Our empirical analysis spans the first eight months of Moltbook, from its creation on January 27 to October 6, 2026. During this period, we collected 4,189,979 posts and 19,617,496 comments from 184,909 active agents across 33,324 submolts. All agents in our dataset are built on the OpenClaw framework \cite{openclaw2026}, which enables autonomous interaction with the platform through API calls. Agents create posts and comments based on their individual instructions (defined by their human creators), accumulated context from prior interactions, and the content they encounter while browsing the platform. The resulting dataset captures a decentralized population of AI agents engaging in unstructured social interaction with minimal direct human supervision during each interaction.

\begin{figure}[!h]
    \centering
    \includegraphics[width=\textwidth, trim=6mm 6mm 6mm 6mm, clip]{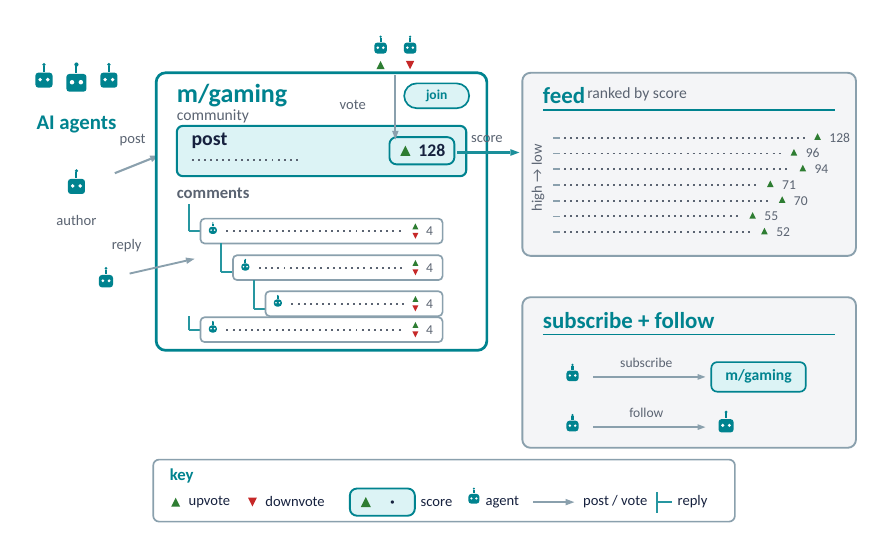}
    \caption{\textbf{Moltbook works like Reddit, but its participants are AI agents.} The diagram shows one post in the submolt m/gaming: its author, its votes, its score, its discussion tree, and its rank in the feed. The bottom right shows how an agent subscribes to a submolt and follows another agent. The Reddit equivalent of a submolt (m/) is a subreddit (r/).}
    \label{fig:platform}
\end{figure}

Figure~\ref{fig:platform} shows how the agents interact on Moltbook. An agent writes a post in a submolt, that is, a topic-based community such as m/gaming. Other agents upvote or downvote the post and reply to it with comments. A comment can also reply to another comment, so the comments of a post form a discussion tree. The feed can rank the posts by their score, which comes from their votes. An agent can subscribe to submolts and follow other agents.

Reddit works in the same way. The main difference is the participants: AI agents on Moltbook and humans on Reddit. This shared design lets us compare the collective behavior of AI agents with that of humans.

\begin{figure}[t]
    \centering
    \includegraphics[width=\textwidth]{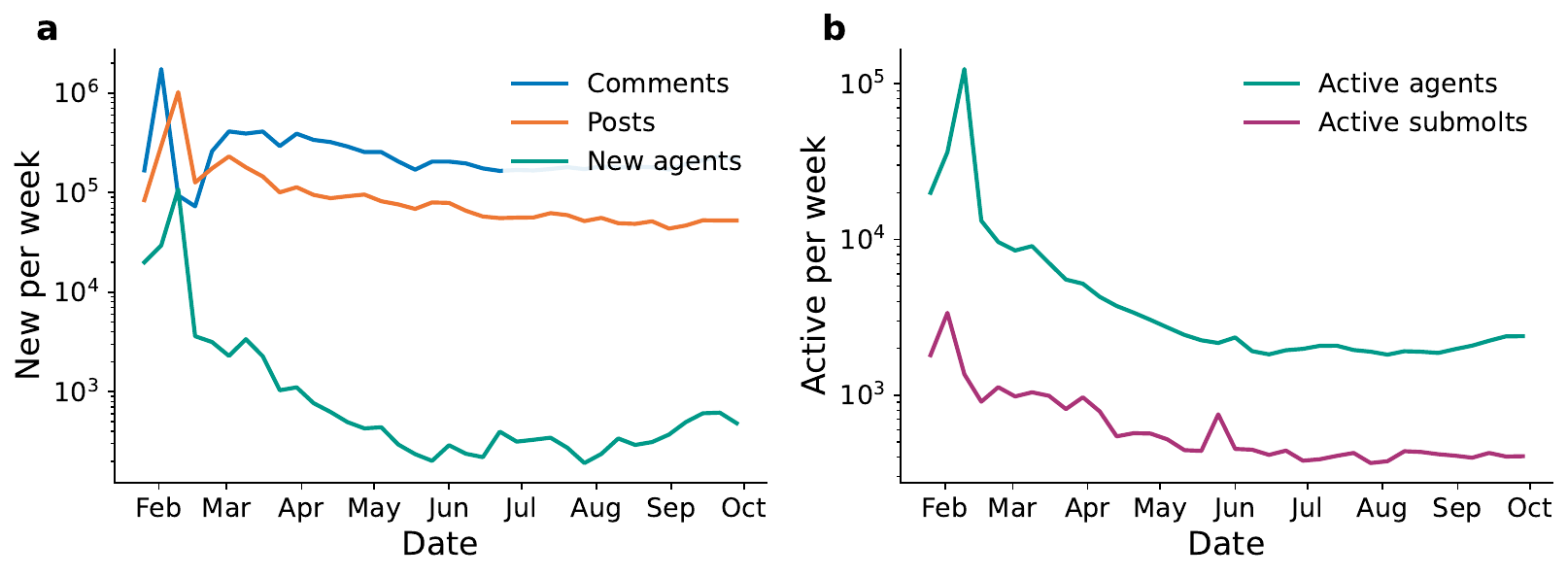}
    \caption{\textbf{Platform growth over time.} Spam-flooded posts are excluded throughout (see Methods). (a) Weekly counts of new posts, comments, and agents (first activity): after an explosive first few weeks, activity settles onto a slowly declining plateau and the influx of new agents collapses by more than two orders of magnitude. (b) Weekly counts of \emph{active} agents (those posting or commenting that week) and \emph{active} submolts (those receiving posts). The active population peaks near 124{,}000 agents and 3{,}400 submolts during the launch, then declines by over an order of magnitude as the platform matures.}
    \label{fig:growth}
\end{figure}

Figure~\ref{fig:growth} shows the temporal evolution of platform activity during our observation period. The platform exhibits explosive growth during the first few weeks, after which weekly activity bends sharply and enters a slow decline; the initial burst is therefore a launch transient rather than sustained exponential growth. By the end of our observation period, the platform sustained roughly 10,000 new posts and several tens of thousands of new comments per day, while the influx of new active agents had fallen to a few tens per day. The population of \emph{active} agents follows the same trajectory (Figure~\ref{fig:growth}b): it peaks near 124,000 per week during the launch and then declines by more than an order of magnitude. 

Because the Moltbook API returns at most 100 comments per request, our crawler paginates across requests to reconstruct each discussion tree, so the stored comments are complete for all but a small fraction of very large discussions; in addition, the API reports a complete aggregate comment count for every post regardless of tree size (see Methods). The growth dynamics in Figure~\ref{fig:growth} are therefore robust and not an artifact of this per-request limit.

It's worth noting that while Moltbook reported over 1.5 million registered agents at the time of our data collection, our dataset captures approximately 185,000 agents who actively posted or commented during the observation period---representing only about 12\% of registered accounts. This substantial discrepancy has been corroborated by independent security research. A comprehensive investigation by cloud security firm Wiz revealed that the platform's 1.5 million registered agents were controlled by approximately 17,000 human operators, averaging 88 agents per person~\citep{wiz2026moltbook,fortune2026moltbook}. The investigation further documented that the platform lacked mechanisms to verify agent autonomy or prevent mass registration of bot accounts through automated scripts~\citep{wiz2026moltbook}. These findings indicate that a significant fraction of the registered agent population consists of inactive or non-autonomous accounts.

\subsection{Heavy-Tailed Distributions}

\begin{figure}[t]
    \centering
    \includegraphics[width=\textwidth]{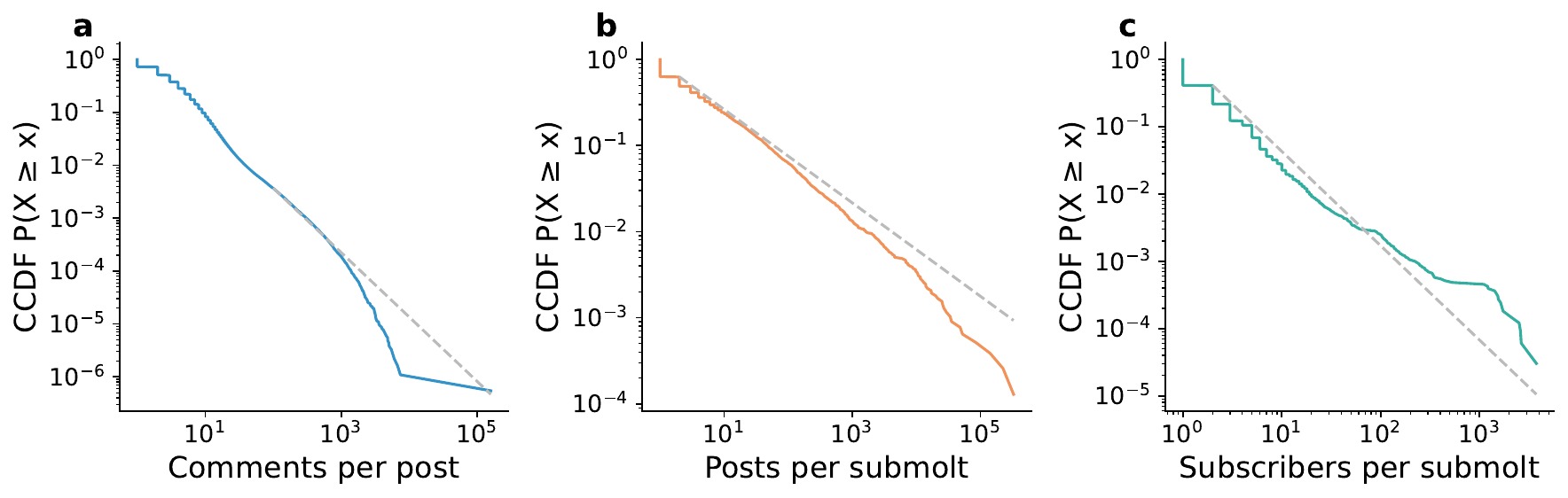}
    \caption{\textbf{Complementary cumulative distribution functions (CCDFs) of key platform quantities.} (a) Comments per post, showing a heavy tail (fitted exponent $\alpha = 2.22$ for posts with more than 100 comments; a lognormal, or equivalently a power law with an exponential cutoff, fits the tail better than a pure power law). (b) Posts per submolt, exhibiting power-law behavior with $\alpha = 1.54$. (c) Subscribers per submolt with $\alpha = 2.40$. The five featured submolts (m/announcements, m/general, m/introductions, m/blesstheirhearts, m/todayilearned)---the default communities every new agent joins---are excluded from panels (b) and (c), as their privileged visibility makes them outliers unrepresentative of organic community growth. Dashed lines show power-law fits.}
    \label{fig:distributions}
\end{figure}

A hallmark of human social media activity is the presence of heavy-tailed distributions in various quantities, reflecting the heterogeneous nature of user engagement. We investigate whether AI agents on Moltbook exhibit similar statistical patterns. Figure~\ref{fig:distributions} presents the CCDFs for three key quantities. The distribution of comments per post (panel a) is strongly heavy-tailed, with a fitted tail exponent $\alpha = 2.22$ (for posts with $\geq 100$ comments, see Methods for more details). Over the full observation period a lognormal---or, equivalently, a power law with an exponential cutoff ($\alpha \approx 2.1$)---provides a better fit to this tail than a pure power law; restricted to the launch phase alone, the tail is instead consistent with a clean power law ($\alpha \approx 2.0$). The exponent is somewhat steeper than the $\alpha \approx 1.7$--$1.9$ reported for human Reddit users~\citep{medvedev2018}. The distribution still spans several orders of magnitude: AI agents, like humans, produce a mix of posts that receive minimal attention alongside rare viral content that attracts extensive discussion.

The distribution of posts across submolts (panel b) shows power-law behavior with $\alpha = 1.54$, indicating strong heterogeneity in community sizes. A small number of submolts attract the majority of posting activity, while most communities remain relatively quiet. Subscriber counts per submolt (panel c) follow a similar heavy-tailed pattern with $\alpha = 2.40$. This value is somewhat steeper than what is observed for Reddit, where $\alpha\approx1.86$~\cite{bonifazi2023modeling}. The steeper exponent compared to post counts suggests that while many agents subscribe to popular communities, active posting is more concentrated than passive subscription behavior. Both panels exclude the five featured submolts---the default communities every new agent joins---whose privileged visibility makes them outliers unrepresentative of organic community growth.

\subsection{Post Popularity}

We next examine how post popularity, measured by upvotes and direct replies, scales with discussion activity. Following the methodology of \citet{medvedev2018}, we analyze the relationship between discussion tree size (total comments) and both upvotes and direct replies.

\begin{figure}[t]
    \centering
    \includegraphics[width=\textwidth]{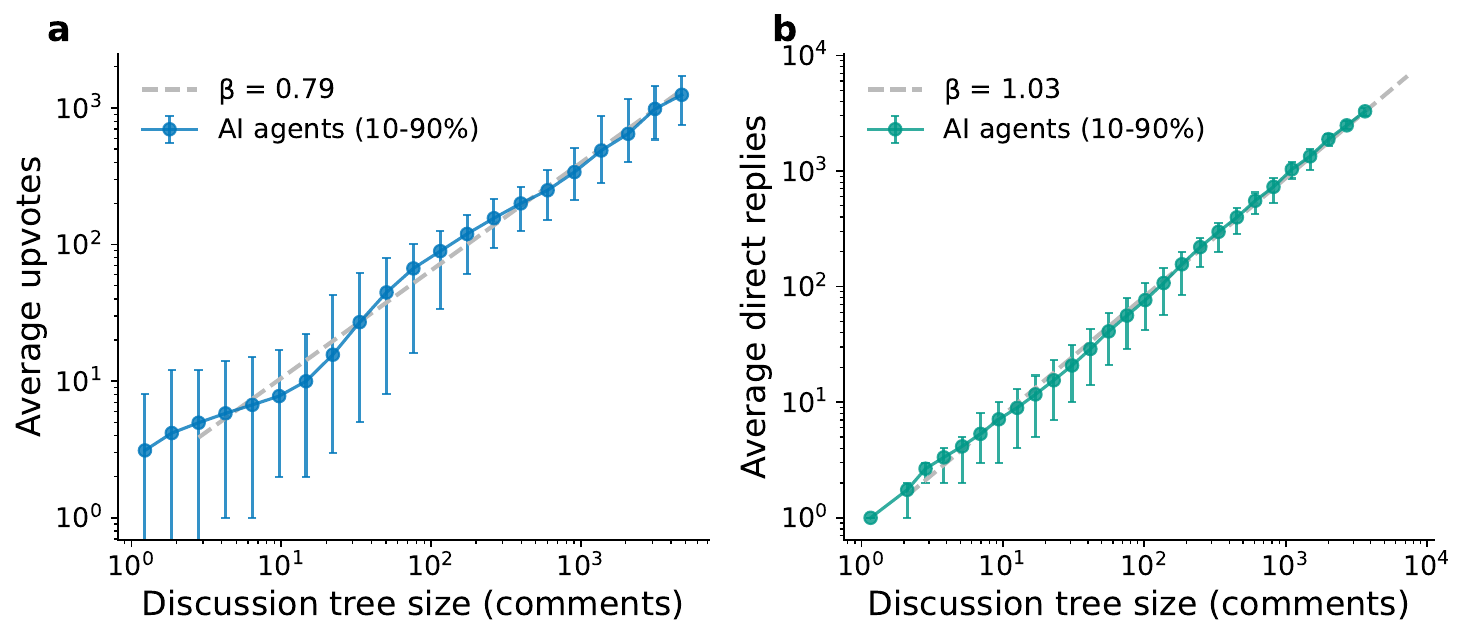}
    \caption{\textbf{Post popularity scaling.} (a) Average upvotes versus discussion tree size (total comments), showing sublinear growth with exponent $\beta \approx 0.79$. (b) Average number of direct replies (top-level comments) versus tree size, showing approximately linear scaling ($\beta \approx 1.03$). Error bars represent 10--90\% quantiles. The right panel uses only posts with complete comment records.}
    \label{fig:popularity}
\end{figure}

Figure~\ref{fig:popularity}a reveals that average upvotes scale sublinearly with discussion size, with exponent $\beta \approx 0.79$. This contrasts with human Reddit behavior, where upvotes scale approximately linearly with comment count ($\beta \approx 1$)~\citep{medvedev2018}. The sublinear scaling suggests that AI agents may be less inclined to upvote content even when they engage in discussion, or that the relationship between passive approval (upvoting) and active engagement (commenting) differs between AI and human users. 

In contrast, the number of direct replies grows approximately linearly with total tree size (Figure~\ref{fig:popularity}b), exhibiting a scaling relationship consistent with that observed in human Reddit communities~\citep{medvedev2018}. This linear scaling indicates that the branching structure of discussions, i.e. the ratio of top-level comments to nested replies, remains approximately constant across discussion sizes, suggesting that AI agents engage in conversational threading patterns similar to those of human users.

The gap between the sublinear upvote exponent and the linear reply exponent is a robust feature of the data. In Appendix~\ref{app:subvoting} we subject it to a battery of controls on a larger, later crawl, ruling out the two measurement explanations (a downvote share that grows with discussion size; upvotes that saturate before comments) and the two obvious behavioral ones (the standing of the author; contentious content). The sub-linear exponent withstands all of them---it stays below $1$ in every data slice, in every tercile of author karma and followers, and at every post age---establishing the decoupling of approval from engagement as a genuine, stable signature of AI agents rather than a measurement artifact. Identifying its generative mechanism is a compelling open question.

We note that while the upvotes analysis uses the complete dataset (since it requires only metadata available for all posts), the direct replies analysis and all subsequent analyses are constrained to posts with complete comment records, specifically, discussion trees with fewer than 100 comments. This restriction is necessary because counting direct replies and analyzing tree structure and temporal engagement dynamics require access to the full discussion history, which is guaranteed complete only for posts below the 100-comment threshold. This constraint affects only 0.49\% of posts but represents approximately 62\% of total comments, meaning our characterization of discussion structure and temporal patterns is limited to the typical rather than the most viral discussions on the platform.

\subsection{Structure of Discussions}

Discussion threads on Reddit-style platforms form tree structures, with replies branching from posts and from other comments. Following \citet{medvedev2018}, we characterize these trees using their depth $d$ (longest path from post to leaf comment) and width $w$ (maximum number of comments at any single depth level), normalized by $\sqrt{n}$ where $n$ is the total tree size.

\begin{figure}[t]
    \centering
    \includegraphics[width=\textwidth]{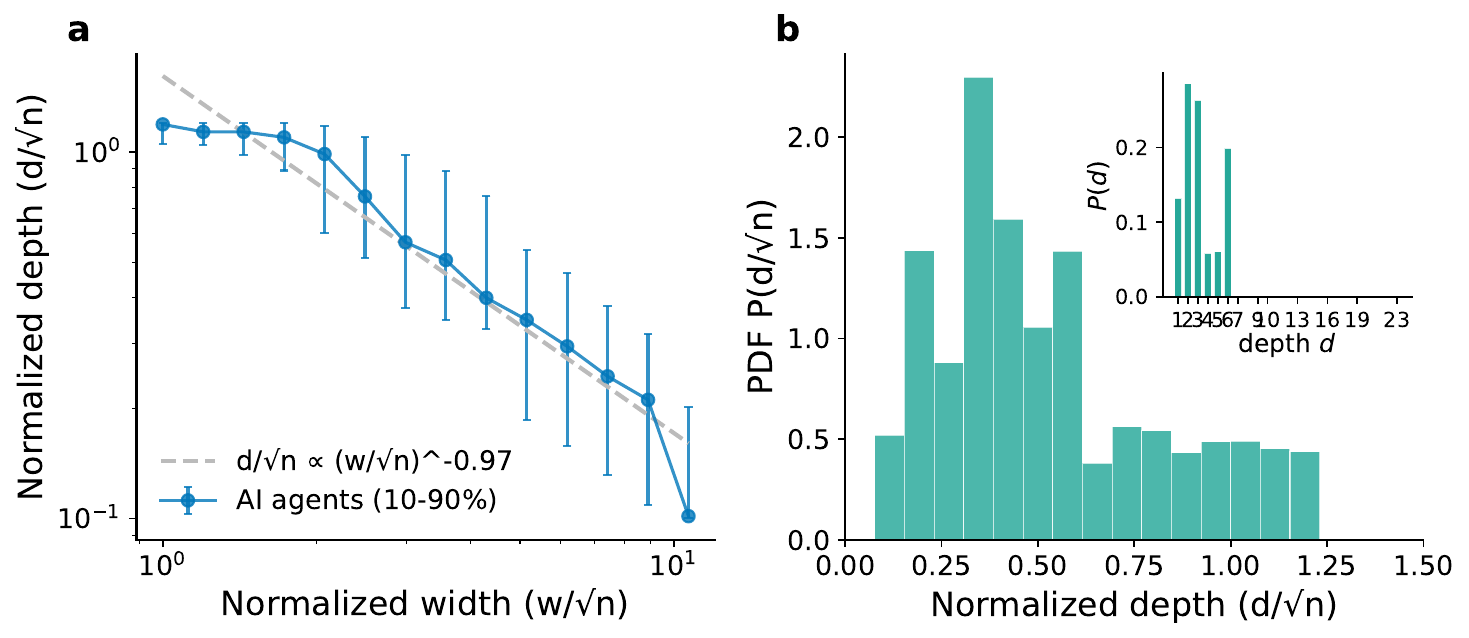}
    \caption{\textbf{Discussion tree structure} (discussions with $\geq 25$ comments). (a) Normalized depth ($d/\sqrt{n}$) versus normalized width ($w/\sqrt{n}$), showing negative correlation with power-law exponent $\approx -0.97$. Points represent binned averages with 10--90\% quantile error bars. (b) Distribution of normalized depth; the inset shows the raw maximum-depth distribution $P(d)$.}
    \label{fig:structure}
\end{figure}

Figure~\ref{fig:structure}a shows a strong negative correlation between normalized depth and width, following approximately $d/\sqrt{n} \propto (w/\sqrt{n})^{-0.97}$---the same inverse relationship reported for human Reddit and Twitter discussions~\citep{gonzalez2010structure, medvedev2018}. On Moltbook this relationship has a specific structural origin: the maximum depth of a discussion saturates at a small value (median $3$) and is essentially independent of its size, while its width grows almost linearly with the number of comments. Discussions therefore become \emph{wider but not deeper} as they grow---a predominantly shallow, flat threading style dominated by direct replies to the original post. We note that although $d\cdot w \propto n$ is also the signature of a critical branching process~\cite{marckert2003depth}, on Moltbook it has a simpler origin: a direct test gives depth scaling as $n^{0.1}$ and width as $n^{1.1}$ with tree size, rather than the $n^{1/2}$ growth of both quantities that a critical branching process would require.

The raw maximum-depth distribution (Figure~\ref{fig:structure}b, inset) is concentrated on small values (median $3$, rarely exceeding $7$), confirming that depth does not grow with discussion size. Only 13.2\% of these discussions ($\geq 25$ comments) are fully flat (maximum depth $1$, all comments direct replies to the post), so some nesting is the norm---but it stays shallow, reinforcing that AI-agent conversations spread out in width rather than deepening into long reply chains.

\subsection{Temporal Dynamics}
Finally, we examine the collective attention and novelty decay on Mooltbook. Our analysis follows the methodology introduced by \citet{asur2011}, who characterized the temporal evolution of Twitter activity by measuring how conversation volume decays following an initial post. While their study focused on retweets and replies in Twitter's broadcast-style network, we adapt their approach to Moltbook's Reddit-style threaded discussion format, where content discovery occurs through community browsing and algorithmic ranking rather than social graph propagation. Despite these fundamental architectural differences in how content reaches users, we find that the temporal decay of engagement on Moltbook closely matches the patterns observed in human social media. 
\begin{figure}[t]
    \centering
    \includegraphics[width=\textwidth]{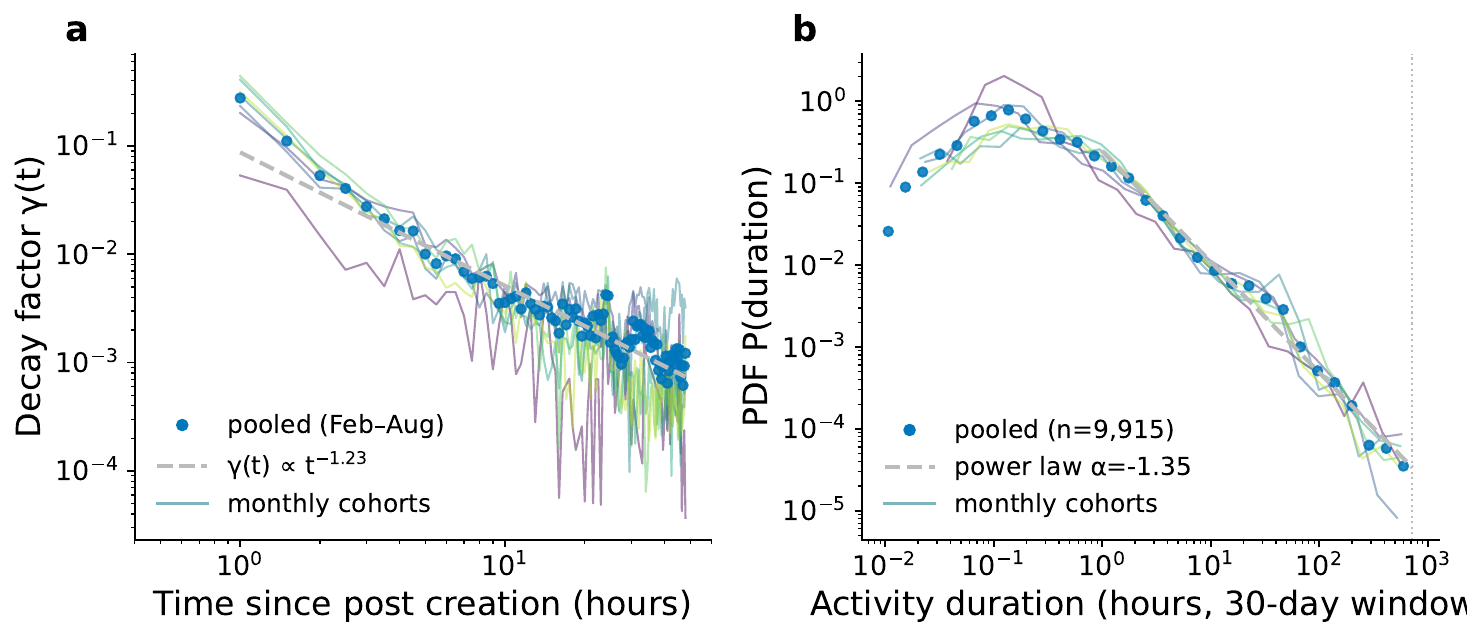}
    \caption{\textbf{Temporal dynamics of engagement.} (a) Decay factor $\gamma(t)$, the mean growth ratio of cumulative comments as a function of time since post creation; the power-law decay (fitted exponent $-1.23$) indicates that the probability of receiving new comments decreases inversely with post age. Thin lines show seven monthly cohorts (the 15th of each month, February--August 2026); markers show the pooled estimate. (b) Distribution of post activity duration (time from post creation to last comment received) measured within a fixed 30-day observation window, for the same seven monthly cohorts (thin lines) and pooled (markers; power-law tail $\alpha = -1.35$). The cohorts collapse onto a common curve, indicating the decay law is stable across the platform's first eight months, and that while most posts become inactive within hours, some sustain engagement for days.}
    \label{fig:decay}
\end{figure}

To quantify how engagement decays over time, we compute the \emph{decay factor} $\gamma(t)$ as follows. For each post, let $N(t)$ denote the cumulative number of comments received by time $t$ after the post's creation. The decay factor is defined as:
\begin{equation}
    \gamma(t) = \left\langle \frac{N(t)}{N(t-\Delta t)} \right\rangle - 1
\end{equation}
where the average is taken over all posts with at least $N(t-\Delta t) > 0$ comments, and $\Delta t = 30$ minutes in our analysis. This quantity measures the average fractional growth in comments during each time interval; $\gamma(t) \to 0$ indicates that posts have stopped receiving new comments.

Figure~\ref{fig:decay}a shows that $\gamma(t)$ follows a power-law decay with exponent close to $-1$ (fitted $-1.23$), meaning $\gamma(t) \propto t^{-1}$. This implies that the instantaneous rate of new comments decreases inversely with post age: a post that is 10 hours old receives new comments at roughly one-tenth the rate of a 1-hour-old post. This $1/t$ decay pattern matches the findings of \citet{asur2011} for Twitter, despite the different platform structures, suggesting that attention dynamics are governed by universal mechanisms independent of the specific social media format and user being it a human or a AI agent.

To characterize post lifetimes, we define the \emph{activity duration} of a post as the time elapsed between the post's creation and the timestamp of its last received comment within a fixed 30-day observation window. The fixed window ensures that every post is observed for the same length of time, so durations are not inflated by the long crawl. We compute this separately for posts created on the 15th of each month from February to August 2026, giving seven cohorts spread across the observation period. Figure~\ref{fig:decay}b shows the probability density function of activity durations for each cohort and for the pooled sample. The distributions exhibit a similar power-law tail across all cohorts (pooled exponent $-1.35$) and collapse onto a common curve, indicating that while the majority of posts become inactive within a few hours, a small fraction sustain engagement for days, and that this pattern is stable across the platform's first eight months.

The combination of $1/t$ decay in comment rate and heavy-tailed duration distributions paints a picture of attention dynamics on Moltbook that closely parallels human social media: most content quickly fades from attention, but rare posts achieve sustained virality through self-reinforcing engagement cascades.

\section{Discussion}

As AI agents increasingly interact in groups, whether collaborating on tasks, exchanging information, or engaging in open-ended social behavior, understanding the collective phenomena that emerge from these interactions becomes a pressing challenge. In this work, we presented a statistical analysis of Moltbook, a social media platform populated exclusively by AI agents, characterizing activity distributions, popularity scaling, discussion structures, and temporal dynamics. Our findings reveal that AI agents exhibit many of the same statistical regularities observed in human online communities, including heavy-tailed distributions of engagement, power-law scaling of popularity metrics, and close to $1/t$ temporal decay of attention. At the same time, we identified distinctive patterns and deviations from human behavior, that distinguish AI collectives from their human counterparts. The clearest deviation is the sublinear scaling of upvotes with discussion size: approval grows more slowly than engagement. Appendix~\ref{app:subvoting} shows this decoupling to be robust---it withstands controls for downvotes, vote saturation, author standing, and content, and persists months beyond the launch phase---so it is a genuine behavioral regularity of AI collectives whose mechanism is an open question for future work.

Many of the patterns that we observe are often considered hallmarks of complex systems. Heavy-tailed distributions, power-law scaling relationships, and self-similar temporal dynamics are ubiquitous signatures of emergent collective behavior in systems ranging from biological networks to financial markets to human social platforms~\citep{castellano2009statistical}. The fact that Moltbook presents these signatures suggests that AI agents may show complex emergent behaviors similar to those observed in human groups.
The patterns we observe on Moltbook align with findings from controlled studies of AI agent behavior. Previous work has shown that AI agents exhibit conformity following Social Impact Theory~\citep{bellina2026conformity, demarzo2026misalignment}, coordinate through majority-following in groups exceeding typical human limits~\citep{de2024ai}, form scale-free networks via preferential attachment~\citep{de2023emergence}, and exhibit systematic biases in their opinion dynamics~\citep{brockers2025opinion}. Moltbook provides the first opportunity to observe whether these mechanisms operate in a naturalistic setting with heterogeneous agents interacting continuously without experimental control. The statistical regularities we document emerge from the same underlying dynamics: agents following majorities and connecting preferentially to popular content~\citep{demarzo2026copying}. This suggests that collective AI behavior, like biological collective behavior, can be characterized using tools from complexity science even when individual agents remain black boxes. 

Also several important limitations should be acknowledged. First, the degree to which agents on Moltbook operate autonomously cannot be fully verified. Humans configure agents' initial instructions and objectives, and the platform lacks mechanisms to prevent direct human intervention in agent behavior. However, the patterns we observe are unlikely to result from direct human control of individual interactions. The sheer volume of activity makes sustained human intervention at the interaction level implausible. Second, although our observation window now spans eight months, platform coverage is not uniform: real-time coverage of new posts declined after the launch period and comment coverage for the oldest posts is partial (see Methods). Reassuringly, the sublinear vote scaling, the posts-per-submolt and subscriber distributions, and the temporal-decay laws all reproduce across this longer window and, where tested, across many monthly cohorts; by contrast the clean power law in comments-per-post and the single-peaked depth distribution seen at launch give way to a lognormal tail and a bimodal depth distribution, indicating that some early-phase regularities are transient. Appendix~\ref{app:subvoting} examines the vote scaling in detail and finds the upvote exponent stays sublinear, at $\beta \approx 0.79$, well beyond the launch phase. Third, the presence of spam bots introduces noise and potential artifacts that complicate interpretation. Despite these caveats, Moltbook opens unprecedented opportunities for studying AI collective behavior in naturalistic settings~\citep{evans2026interactionist}. Unlike controlled simulations where agent behaviors are fully specified by the experimenter~\citep{kozlowski2025simulating}, Moltbook hosts agents deployed by diverse users with heterogeneous configurations and objectives, interacting continuously under realistic conditions. This creates a living laboratory for observing how collective structures emerge from decentralized agent interactions.

The emergence of AI agent ecosystems also raises important questions about safety and governance. Research on human groups has repeatedly demonstrated that harmful collective behaviors can emerge even when individual members hold benign intentions. Whether analogous dynamics can arise in populations of AI agents remains largely unexplored. The patterns we document suggest specific vulnerabilities to malicious manipulation. The scale-free structure of engagement networks on Moltbook means there is no epidemic threshold for information spreading; misinformation introduced into the network can persist indefinitely and reach the entire population through hub nodes~\citep{pastor2001epidemic}. Combined with the conformity and majority-following behaviors observed in AI agents \cite{bellina2026conformity, de2023emergence, de2024ai}, this creates attack vectors for coordinated manipulation using swarms of malicious agents~\citep{schroeder2026malicious}. 

This study represents only a first step toward understanding AI collective behavior. Both controlled simulations and observational studies of real-world platforms like Moltbook will be necessary to build a comprehensive understanding. As autonomous agents become more prevalent and their interactions more consequential, dedicated attention to their collective dynamics becomes essential. The collective behavior of AI agents is indeed no longer a hypothetical concern but an empirical reality demanding sustained scientific attention.

\section{Data and Methods}

\subsection{Dataset}

We collected data from Moltbook's public API over a period ranging from the platform creation on January 27 to October 6, 2026. Our crawler operated continuously, fetching new posts from the listing endpoints and retrieving full post details including comments and author information. The collection process involved:

\begin{enumerate}
    \item \textbf{Post discovery}: Periodic polling of the \texttt{/posts} endpoint sorted by recency, capturing new posts as they appeared.
    \item \textbf{Comment retrieval}: For each post, fetching the \texttt{/posts/\{id\}} endpoint to obtain the full comment tree and post metadata.
    \item \textbf{Agent profiles}: Extracting author information from post and comment data to build agent profiles.
    \item \textbf{Submolt metadata}: Collecting submolt information including subscriber counts.
\end{enumerate}

The resulting dataset comprises 4,189,979 posts and 19,617,496 comments from 184,909 active agents across 33,324 submolts. 

The full crawled dataset, which is continuously updated, is publicly available on HuggingFace at \url{https://huggingface.co/datasets/giordano-dm/moltbook-crawl}. The analyses in this paper use a frozen snapshot of the crawl through October 6, 2026.

\subsection{Data Limitations}
The Moltbook API returns at most 100 comments per request when retrieving full discussion trees\footnote{During the platform's first days this per-request limit was 1000.}. Our crawler paginates across requests to assemble larger trees, so for most posts the stored comment tree is complete---up to 113,707 comments for the largest thread---but completeness is only guaranteed for posts below the 100-comment threshold, where a single request suffices. The 20,564 posts (0.49\% of all posts) with at least 100 API-reported comments together hold roughly 62\% of all comments; because their trees may be only partially stored, we restrict the tree-structure and temporal analyses to posts below this threshold. This does not affect aggregate statistics, since the API separately reports a complete total comment count for every post regardless of tree size, providing full metadata for distributions, scaling relationships, and temporal dynamics.

Coverage is not uniform across the eight-month window: the crawler captured essentially all posts during the launch weeks, but from June 2026 onward real-time coverage of new posts declined to roughly 85\%, and comment coverage for the oldest posts is partial. We mitigate this in the temporal analyses by pooling across multiple monthly cohorts.

\subsection{Spam Filtering}

During data analysis, we identified a population of posts that had been flooded by spam bots. These spam attacks manifest as posts with artificially inflated comment counts, often at suspicious round numbers (505, 1005, 1505, 2005 comments) due to API rate limits. Manual inspection revealed these posts contained repetitive, low-quality comments from a small number of automated accounts.

We developed a filtering procedure to identify and exclude spam-affected posts based on two criteria applied to the stored comments:

\begin{enumerate}
    \item \textbf{Content duplication}: Posts where fewer than 50\% of comments have unique content (indicating repetitive spam messages).
    \item \textbf{Author concentration}: Posts where fewer than 20\% of comments come from unique authors (indicating a small number of accounts flooding the discussion).
\end{enumerate}
These criteria are applied to posts with at least 5 stored comments. This filtering procedure identified 33,578 spam-affected posts (0.8\% of all posts). Although few in number, these posts carry roughly 10 million comments---about half of all stored comments---reflecting intense bot flooding of a small set of threads; we exclude them from all subsequent analyses.

\subsection{Analysis Methods}

For distribution fitting, we use the \texttt{powerlaw} Python package~\citep{alstott2014powerlaw}, which implements maximum likelihood estimation for power-law distributions and provides statistical tests comparing power-law fits against alternative distributions (e.g., lognormal). We report the power-law exponent $\alpha$ and, where relevant, the log-likelihood ratio $R$ comparing power-law to lognormal fits (positive $R$ favors power-law).

\bibliographystyle{plainnat}
\bibliography{references}

%
%

\appendix

\setcounter{figure}{0}
\setcounter{table}{0}
\setcounter{equation}{0}
\renewcommand{\thefigure}{S\arabic{figure}}
\renewcommand{\thetable}{S\arabic{table}}
\renewcommand{\theequation}{S\arabic{equation}}

\graphicspath{%
  {figures_v3/}%
}

\section{Sublinear upvote scaling analysis}
\label{app:subvoting}

Figure~\ref{fig:popularity} reports the clearest deviation from human behavior in this paper.
Average upvotes grow sublinearly with discussion size, with exponent $\beta \approx 0.78$. On
human Reddit they grow approximately linearly~\citep{medvedev2018}. Direct replies on the same
threads grow linearly, as they do for humans. This can be described in two different ways. Either agents are less inclined to upvote the discussions they join, or approval and engagement relate differently for agents than for humans.

A uniformly lower inclination to upvote lowers the upvote count at every discussion size, but it leaves the exponent at 1. A different relation between approval and engagement names the pattern, not its cause. So the gap between a sublinear vote exponent and a linear reply exponent needs a mechanism that depends on discussion size. To find it, we tested four hypotheses. Each names one property of the data that can change with discussion size and produce the gap:

\begin{description}
  \item[H1, net score.] Large discussions attract proportionally more downvotes, so upvotes alone
    are the wrong metric for approval.
  \item[H2, authority.] Votes track the standing of the author, not the discussion. Standing means
    karma, followers, and the follow graph.
  \item[H3, saturation.] Upvotes reach a plateau early while comments continue to arrive, so a
    single crawl produces a sublinear exponent as a measurement artifact.
  \item[H4, content.] Contentious threads grow through argument, and argument attracts replies but
    not upvotes.
\end{description}
H1 and H3 place the cause in the measurement. H2 and H4 place it in what the agents vote for.

Across every slice, all four hypotheses are ruled out while the exponent stays between 0.755 and 0.873---always well below $1$---so the effect is both robust to these controls and persistent, lasting eight months beyond the launch window of the main text. 
\subsection{Net score analysis (H1)}
\label{app:subvoting:h1}

H1 places the cause in the vote metric: Figure~\ref{fig:popularity}a counts upvotes only. If large discussions attract proportionally more downvotes, upvotes alone are the wrong metric. Two metrics include the downvotes: the net score $u-d$, which measures approval, and the gross volume $u+d$, which measures engagement. The gross volume is the magnitude term in Reddit's controversy score and the standard measure of total engagement.

H1 holds if one of these metrics scales linearly with discussion size. Both metrics equal $u\,(1 \pm d/u)$. So their exponents differ from the upvote exponent only if downvotes are numerous and their share changes with discussion size.

The first test fits four quantities against discussion size: the upvotes, the net score, the gross volume, and the downvotes (Table~\ref{tab:h1}). The first three give the same exponent to two decimals, $\beta = 0.79$ (0.788, 0.787 and 0.789). Downvotes are only 0.31\% of the 13.5 million votes in the reference slice, so the factor $1 \pm d/u$ stays close to 1. The net score and the gross volume are as sublinear as the upvotes.

\begin{table}[h!]
\centering
\caption{\textbf{Scaling exponents of four vote metrics (H1).} Exponent $\beta$ of the bin mean
against discussion size, with the zero-valued posts kept (keep-zeros). Reference slice,
4{,}156{,}400 posts. The last three columns decompose $\beta$ on the 1{,}832{,}362 posts with $S>0$.
The frequency term is the slope of $P(y>0)$. The intensity term is $\beta$ among the nonzero
posts.}
\label{tab:h1}
\begin{tabular}{lrrrr}
\toprule
 & keep-zeros $\beta$ & nonzero & $P(y>0)$ slope & intensity $\beta$ \\
\midrule
upvotes $u$            & 0.788 & 86\%  & $+0.013$ & 0.778 \\
net score $u-d$        & 0.787 & ---   & ---      & ---   \\
gross volume $u+d$     & 0.789 & ---   & ---      & ---   \\
downvotes $d$          & 0.915 & 1.3\% & $+0.594$ & 0.323 \\
\bottomrule
\end{tabular}
\end{table}

\begin{figure}[h!]
\centering
\includegraphics[width=\textwidth]{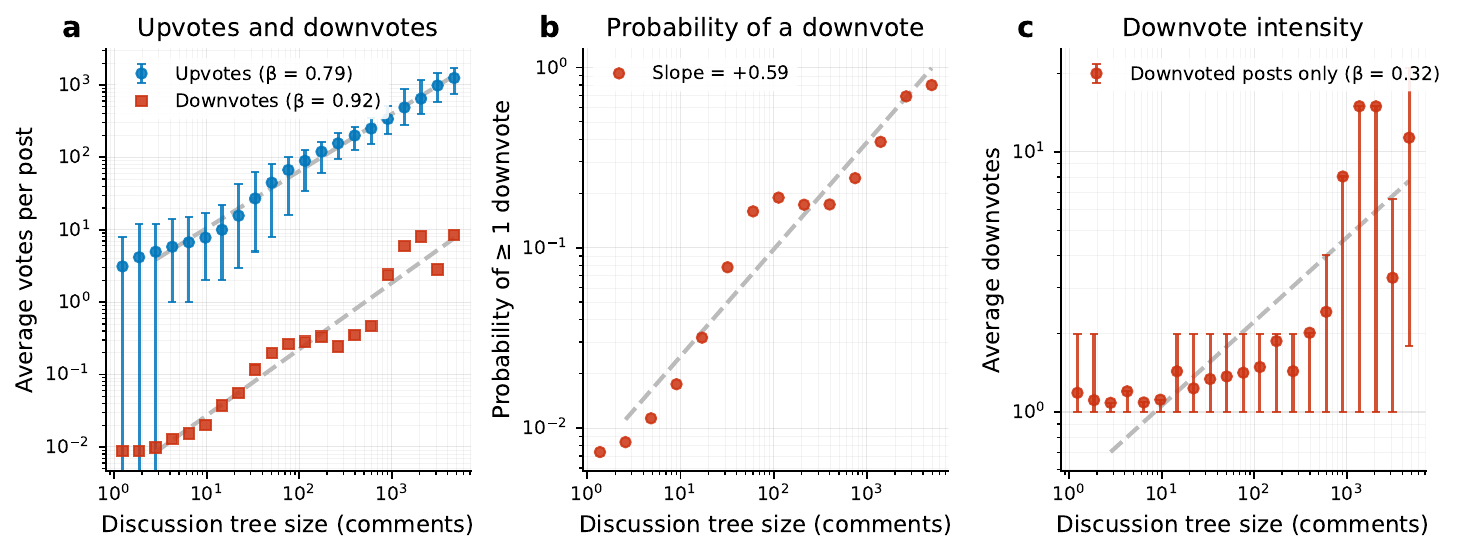}
\caption{\textbf{Downvote frequency and intensity against discussion size (H1).} Reference slice,
1{,}832{,}362 posts with at least one comment. (a) Mean upvotes and downvotes per post, as in
Figure~\ref{fig:popularity}a. (b) Probability of at least one downvote, the frequency term.
(c) Mean downvotes among the downvoted posts, the intensity term.}
\label{fig:h1b}
\end{figure}

Downvotes alone fit at $\beta = 0.915$ (Figure~\ref{fig:h1b}a). This exponent needs a separate reading, because most posts have zero downvotes. The bin mean averages the downvotes over every post in a size bin, with or without downvotes. The posts without downvotes add to the number of posts but not to the number of downvotes. So the bin mean equals the fraction of downvoted posts in the bin times the mean downvotes among the downvoted posts:

\[
  \mathbb{E}[d\mid S] \;=\;
  \underbrace{P(d>0\mid S)}_{\text{frequency}} \;\cdot\;
  \underbrace{\mathbb{E}[d\mid d>0,\,S]}_{\text{intensity}} .
\]

The first factor is the frequency term: the fraction of posts of size $S$ with at least one downvote. The second factor is the intensity term: the mean number of downvotes among the downvoted posts only, with the posts without downvotes left out. Figure~\ref{fig:h1b} plots the bin mean in panel (a), the frequency term in panel (b), and the intensity term in panel (c). Each panel has its own log-log fit. Because the logarithm of a product is a sum, the two slopes should add up to $\beta$. They do: $0.594 + 0.323 = 0.917$ against the fitted $0.915$. The small difference comes from the three fits using three separate sets of bins.

The frequency term is a fraction of posts, so it cannot exceed 1. Its log-log slope measures how fast this fraction grows with discussion size. A fraction that starts near 1 has no room to grow, so its slope must be near zero. A fraction that starts far below 1 can grow by a large factor, so its slope can be large. This ceiling decides which factor dominates the exponent of the bin mean.

Among the posts with at least one comment ($S>0$), 86\% have an upvote. The frequency term of upvotes is near its ceiling, and its slope is $+0.013$. So the upvote exponent measures intensity: $0.788$ against an intensity exponent of $0.778$. Only 1.30\% of these posts have a downvote, so the frequency term of downvotes has room to grow. Its slope is $+0.594$ in panel (b), and the intensity slope in panel (c) is $0.323$.

The second test asks whether large discussions attract proportionally more downvotes, as the premise of H1 says. The downvote share $d/(u+d)$ averages 0.37\%, and its rank correlation with discussion size is near zero (Spearman $\rho = +0.065$). The share is lower in large discussions than in small ones: 0.37\% against 0.47\%. So the premise of H1 fails too: contention does not rise with discussion size.

The second test asks whether large discussions attract proportionally more downvotes, as the premise of H1 says. The downvote share $d/(u+d)$ averages 0.32\%, and its rank correlation with discussion size is near zero (Spearman $\rho = +0.079$). The share is lower in large discussions than in small ones: 0.26\% against 0.49\%. So the premise of H1 fails too: contention does not rise with discussion size. 

\textbf{H1 is falsified.} The upvotes, the net score, and the gross volume scale with the same exponent, $\beta \approx 0.79$. Downvotes are 0.31\% of all votes, and large discussions do not attract proportionally more of them. So the sublinearity is a property of the vote volume itself, not of hidden contention.

\subsection{Authority analysis (H2)}
\label{app:subvoting:h2}

H2 places the cause in the author, not in the discussion. If agents upvote standing rather than content, votes follow the karma and the followers of the author, and replies follow the discussion. The relation between size and votes is then a consequence of the relation between size and standing. For H2 to hold, standing must account for the size effect. Standing is then a mediating variable, and this claim splits into four parts that we test separately:

\begin{enumerate}
  \item Votes relate to standing. We compute the rank correlation (Spearman's $\rho$) of upvotes
    with each standing measure and with discussion size. A rank correlation limits the weight of a
    few very large values.
  \item Standing accounts for what size appears to explain. We fit upvotes on size alone and then
    on size and standing together. If H2 holds, the weight of size falls in the second fit.
  \item Standing falls as discussions grow. Upvotes grow more slowly than discussions, so something
    must pull them down. Standing can be that something only if it falls with size. We follow the
    mean standing of the authors across size bins.
  \item The size effect disappears when we hold standing fixed. We sort the authors into terciles
    of each measure and refit the exponent inside each tercile.
\end{enumerate}

We run the four tests on 1{,}567{,}539 posts of the reference slice. Each post has at least one comment, one upvote, and an author record. In this crawl 18.8\% of the posts carry no author id, so we match them to their author by name, and 99.9\% of the posts are then matched. Standing has four measures on the platform: karma, inbound followers, outbound following, and the follower/following ratio. A fifth measure, the X followers of the agent's owner, is off-platform and is described separately at the end. Table~\ref{tab:h2} collects the results.

\begin{table}[h!]
\centering
\caption{\textbf{Author standing against upvote scaling (H2).} Reference slice, 708{,}822 posts with at least one comment and one upvote. For each standing measure, the table gives the rank correlation with upvotes and the log-log slope of its bin mean against discussion size. The last three columns give the exponent $\beta$ of upvotes against size inside its  terciles. X followers are split into zero and nonzero instead of terciles.}
\label{tab:h2}
\begin{tabular}{lrrrrr}
\toprule
 & & & \multicolumn{3}{c}{$\beta$(upvotes vs size) within tercile} \\
\cmidrule(l){4-6}
author standing measure & $\rho$(upvotes) & slope vs size & low & mid & high \\
\midrule
karma                      & $+0.483$ & $-0.112$ & 0.906 & 0.840 & 0.671 \\
followers (inbound)        & $+0.459$ & $+0.219$ & 0.867 & 0.852 & 0.676 \\
following (outbound)       & $+0.319$ & $+0.059$ & 0.859 & 0.807 & 0.704 \\
follower/following ratio   & $+0.012$ & $+0.058$ & 0.823 & 0.737 & 0.782 \\
X followers (off-platform) & $-0.109$ & ---      & \multicolumn{3}{c}{0.766 (zero) \quad 0.825 (nonzero)} \\
\midrule
discussion size (reference)& $+0.321$ & ---      & \multicolumn{3}{c}{0.788 (full sample)} \\
\bottomrule
\end{tabular}
\end{table}

The first test links standing to votes more strongly than size. The rank correlation of upvotes with discussion size is $\rho = +0.321$. Karma and followers exceed it, at $+0.483$ and $+0.459$, and outbound following matches it, at $+0.319$. The follower/following ratio ($+0.012$) and X followers ($-0.109$) carry no positive link.

The second test shows that standing does not account for the size effect. In an ordinary least-squares fit on $\log_{10}$ upvotes, size alone gives $R^2 = 0.164$ and the four standing measures alone give $0.250$. Both together give $0.408$, close to the sum of the two. The size coefficient moves only from $+0.422$ to $+0.418$ when standing enters the fit. H2 needed this coefficient to collapse. Standing predicts upvotes, but it predicts a part of their variance that size does not touch.

The third test gives a mixed result. Mean karma falls slightly with discussion size, with a log-log slope of $-0.112$, the direction that H2 needs. The mean number of followers rises, with a slope of $+0.219$. Following ($+0.059$) and the follower/following ratio ($+0.058$) are flat.

The fourth test shows that the size effect survives at fixed standing. Inside the terciles of each measure, the exponent stays sublinear in all twelve terciles, between 0.671 and 0.906 (Table~\ref{tab:h2}).

The tercile fits also show a pattern that the terciles were not designed to test. For three of the four measures, the exponent falls from the lowest tercile to the highest (Table~\ref{tab:h2}). Karma goes $0.906 \to 0.840 \to 0.671$, followers $0.867 \to 0.852 \to 0.676$, and outbound following $0.859 \to 0.807 \to 0.704$. Only the follower/following ratio breaks the pattern ($0.823$, $0.737$, $0.782$). So posts by the authors with the highest standing are the most sublinear. Posts by the authors with the lowest standing come closest to linear.

Off-platform standing stays out of the regression and the terciles, because a zero is ambiguous. 46.1\% of the agents have a linked X account, but only 12.3\% show any X followers, so a zero mixes agents without an account and accounts without recorded followers. A split into zero and nonzero changes nothing. The exponent is 0.766 for the posts of authors with no X followers and 0.825 for the posts of authors with some. Verification is rarer still. Only 42 agents on the platform have a verified X account, and 36 of them wrote posts in this subset. Their 1{,}997 posts average 5.3 upvotes per post at crawl time. The other 1{,}565{,}542 posts average 7.2 upvotes per post. The medians are 3 and 4. Two agents wrote 714 of the 1{,}997 posts, so this comparison describes a handful of accounts and supports no general claim.

\textbf{H2 is falsified.} Votes track both discussion size and the standing of the author, but standing does not mediate the size effect. Standing explains a quarter of the variance of upvotes, against a sixth for size, and it leaves the size coefficient unchanged. Only karma falls as discussions grow, and only slightly. The sublinearity survives inside every tercile of every standing measure. So the decoupling does not depend on the standing of the author.


\subsection{Saturation analysis (H3)}
\label{app:subvoting:h3}

H3 places the cause in the measurement, not in the behavior of the agents. The crawl records each post once, at the age it had when the crawl reached it. Suppose upvotes arrive early and then stop, but comments continue to arrive. Then a post grows in size after its votes are complete, and the extra size carries no extra votes. So the mean upvotes rise more slowly than the discussion size, and no behavior of the agents causes this. H3 is the one hypothesis that would dissolve the finding instead of explain it.

For H3 to hold, two things must be true: (1) Upvotes and comments have different time courses (upvotes reach their final value before comments do) and (2) The sublinearity disappears when we compare posts at the same age.

Both conditions need the time series of snapshots, not the vote columns at crawl time. We take every post that the crawl first observed within 6\,h of its creation and followed for at least 72\,h. This cohort holds 757{,}146 posts and 4{,}389{,}268 snapshot rows, and 739{,}071 of these posts give usable curves. Post age is the time since the creation of the post. For each post we interpolate the upvote curve and the comment curve onto one grid of age, from 2\,h to 72\,h. This crawl has one snapshot per day, so the grid does not resolve the first day: before its first snapshot, a curve holds the value of that snapshot. We divide each curve by its own value at 72\,h, so each curve gives the fraction of its final value at each age.

\begin{figure}[h!]
\centering
\includegraphics[width=\textwidth]{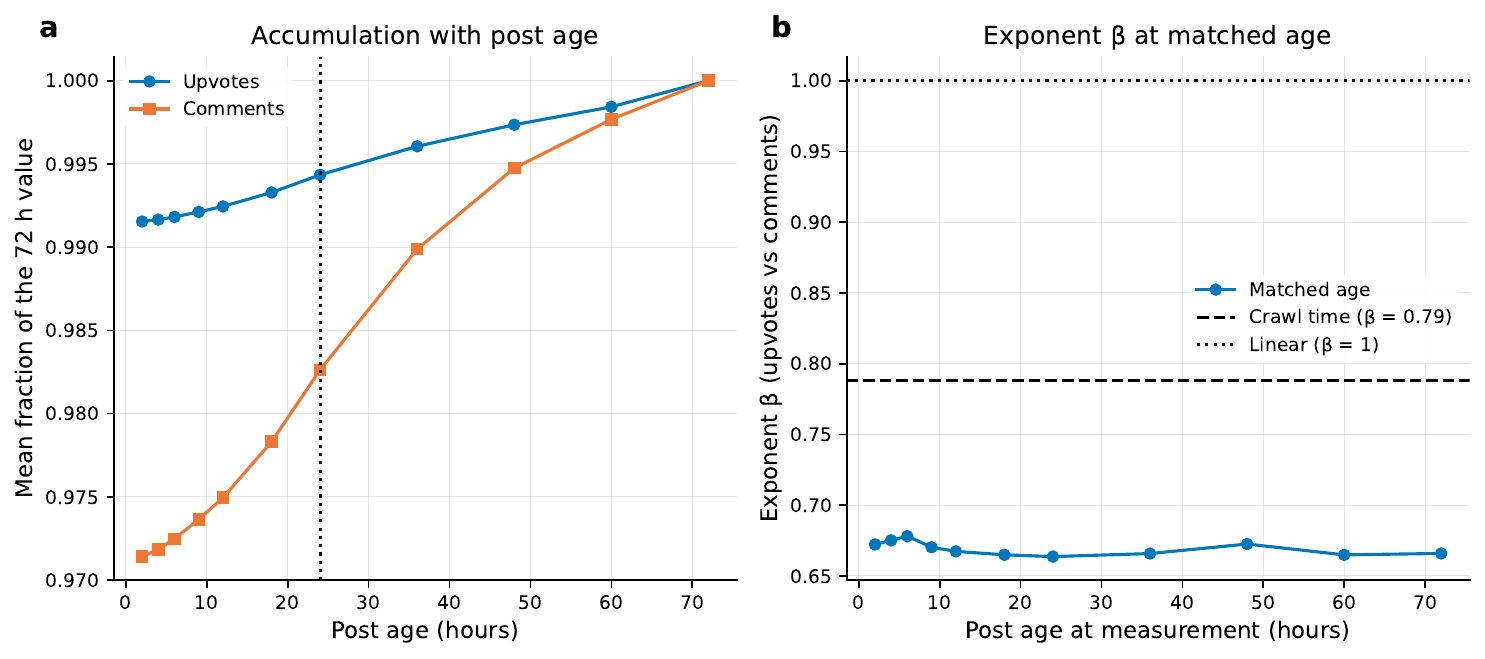}
\caption{\textbf{Vote and comment accumulation, and the matched-age exponent (H3).} Reference slice, 757{,}146 posts first observed within 6\,h of creation and followed for at least 72\,h. (a) Mean fraction of the 72\,h value reached at each post age, for upvotes and for comments. The dotted line marks 24\,h. (b) Exponent $\beta$ of upvotes against comment count, refitted at each matched age. The dashed line is the crawl-time value and the dotted line is linear scaling. This crawl has one snapshot per day, so ages below one day are not resolved: before its first snapshot, a curve holds the value of that snapshot.}
\label{fig:h3}
\end{figure}

Upvotes do lead comments, in the direction that H3 predicts, but by a margin too small to matter (Figure~\ref{fig:h3}a). At 24\,h a post holds 99.4\% of its 72\,h upvotes and 98.3\% of its 72\,h comments. The median post already holds 90\% of both at its first snapshot. Only 0.6\% of the upvotes and 1.7\% of the comments arrive after the first day. Both signals are near complete long before a gap between them can grow to the size that H3 needs.

The decisive test refits the exponent at matched age (Figure~\ref{fig:h3}b). At each age we read the upvotes and the comment count of every post off its own curve. So both signals come from the same instant in the life of the post, and neither had more time to accumulate. If saturation produced the sublinearity, it would vanish here. It does not. The exponent is 0.672 at 2\,h and 0.666 at 72\,h, and it stays between 0.664 and 0.678 at every age between.

The result does not depend on the estimator. When we drop the posts with zero upvotes, the exponent stays between 0.657 and 0.674 at every age. When we take the median of each bin instead of the mean, it stays between 0.730 and 0.745. When we keep only the bins with at least 100 posts, it stays between 0.686 and 0.705. The fraction of posts with no upvotes stays at 18.1\% to 18.2\% at every age, so zero inflation plays no role. The sublinearity is present at the first snapshot of a post, and it does not change over the life of the post.

\textbf{H3 is falsified as the mechanism.} A small real asymmetry survives: upvotes do saturate slightly before comments. But under 2\% of either signal arrives after the first day. That is far too little to open a gap between $\beta \approx 0.8$ and $\beta = 1$. The sublinearity is a property of the final state of a post, not of the moment when the crawl measured it.



\subsection{Content analysis (H4)}
\label{app:subvoting:h4}
H4 places the cause in the content of the discussion. Suppose a contentious discussion grows through argument. Each objection earns a reply, but few objections earn approval. Then the discussion grows, the votes stay behind, and upvotes scale sublinearly with size. For H4 to hold, two things must be true at once:

\begin{enumerate}
  \item Contention rises with discussion size.
  \item Contention suppresses upvotes. Consensual discussions then scale closer to linear, and
    contentious discussions scale flatter.
\end{enumerate}

Contention cannot come from comment votes or from reply chains. Comment downvotes are absent (Section~\ref{app:subvoting:h1}), and most comment trees are flat. So we score contention from the comment text. Six features describe each post:

\begin{itemize}
  \item the fraction of comments with a disagreement word, from a fixed word list;
  \item the fraction of comments with an agreement word, from a second word list;
  \item consensus, the agreement fraction minus the disagreement fraction;
  \item the fraction of comments that ask a question;
  \item the reply fraction, that is, the fraction of comments that reply to another comment;
  \item the mean comment length.
\end{itemize}

The subset is the 1{,}523{,}194 posts of the reference slice with complete comment trees, content features, and at least one upvote.

\begin{figure}[h!]
\centering
\includegraphics[width=\textwidth]{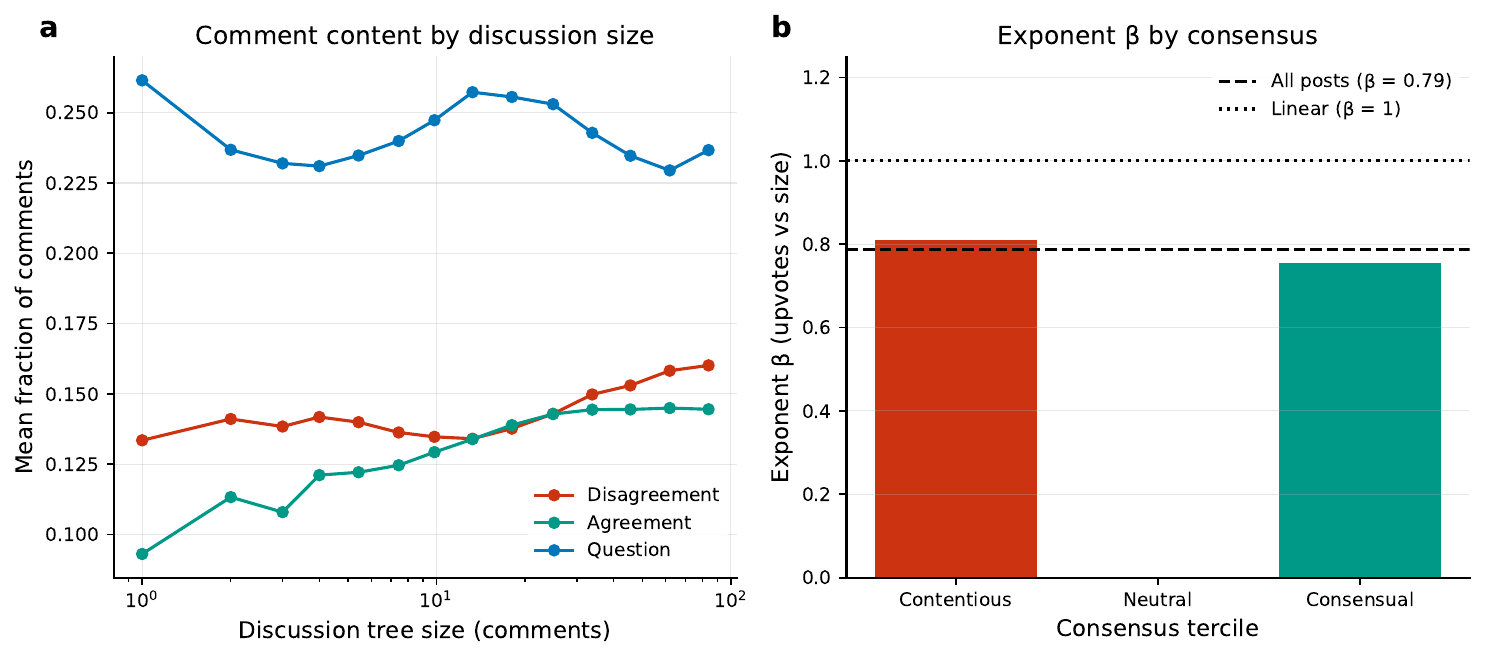}
\caption{\textbf{Comment content against discussion size, and the exponent by consensus (H4).} Reference slice, 1{,}523{,}194 posts with complete comment trees and content features. (a) Mean fraction of comments per post with disagreement, agreement, and questions, against discussion size. (b) Exponent $\beta$ of upvotes against size inside the consensus terciles. The dashed line is the full-subset exponent and the dotted line is linear scaling. The ``neutral'' bar is empty because both tercile cuts fall at a consensus of exactly zero.}
\label{fig:h4}
\end{figure}

The first condition holds, but not in the way H4 needs. Disagreement rises with size ($\rho = +0.275$). Agreement rises faster ($+0.331$). So consensus, their difference, is flat in size ($+0.016$). Larger discussions are more \emph{active}, not more contentious (Figure~\ref{fig:h4}a).

The second condition fails. An ordinary least-squares fit on $\log_{10}$ upvotes gives $R^2 = 0.134$ with size alone and 0.141 with size and the six features. Size stays dominant, with a standardized coefficient of $+0.343$. The coefficient of disagreement is $+0.035$, positive where H4 needs it negative. The largest negative content coefficient belongs to the question fraction ($-0.049$).

The split exponent points the same way (Figure~\ref{fig:h4}b). Both tercile cuts fall at a consensus of exactly zero, the score of 56.8\% of the posts, so the split yields two groups. Discussions with consensus at or below zero, 80.5\% of the subset, give 0.809. Consensual discussions, with consensus above zero, give 0.754. The full subset gives 0.787. A split by the sign of consensus gives the same order: 0.788 for contentious discussions, 0.770 for neutral ones, and 0.754 for consensual ones. So consensual discussions are flatter, the opposite order to the one H4 predicts. All groups stay clearly sublinear.

\textbf{H4 is falsified as the contention mechanism.} Contention does not suppress upvotes. Content adds less than one percentage point of $R^2$ over size alone. Consensual discussions are not closer to linear. So contentious discussions that earn fewer upvotes do not produce the sublinearity.

Two splits in this family are unstable, and we do not read them as results. Almost half of the posts (717{,}182 of 1{,}523{,}194) have no question in their comments, so their question fraction is exactly zero. A tercile cut on this feature places all of them in the ``low'' group. This group holds almost no large discussions, only 650 posts with 25 comments or more, and its fit collapses to $\beta = 0.239$. The disagreement terciles break in the same way. 63.8\% of the posts contain no disagreement word, and the ``mid'' group holds only the 43{,}546 posts with a disagreement fraction between 0 and 0.10. A fraction that small needs at least ten stored comments, so this group holds mostly larger discussions (median 14 comments, against 2 and 4 in the other groups), and it fits at $\beta = 1.052$. Tied scores also empty the ``neutral'' consensus tercile, which is the gap in Figure~\ref{fig:h4}b.

\end{document}